\documentclass[a4wide]{article}

\date{\today}

\newcommand{\insertplot}[5]{\begin{figure}
 \hfill\hbox to 0.05in{\vbox to #5in{\vfill
 \inputplot{#1}{#4}{#5}}\hfill}
 \hfill\vspace{-.1in}
 \caption{#2}\label{#3}
 \end{figure}}
 \newcommand{\inputplot}[3]{
 \special{ps: plotfile #1}
}
\newcounter{fig}   \newcommand{\lbfig}[1]{\refstepcounter{fig}
\label{#1} }

\usepackage{epsfig}
\usepackage{amsmath}
\usepackage{amsfonts}
\usepackage{graphicx}
\usepackage[german, english]{babel}
\usepackage{amsmath}
\usepackage{amssymb}
\usepackage{ifthen}
\usepackage{epsfig}

\begin{document}
\newcommand{\BA}{\ensuremath{\mathbf{A}}}
\newcommand{\Beta}{\ensuremath{\mathbf{\eta}}}
\newcommand{\BV}{\ensuremath{\mathbf{V}}}
\newcommand{\Bn}{\ensuremath{\mathbf{n}}}
\newcommand{\BW}{\ensuremath{\mathbf{W}}}
\newcommand{\BD}{\ensuremath{\mathbf{D}}}
\newcommand{\BP}{\ensuremath{\mathbf{P}}}
\newcommand{\BU}{\ensuremath{\mathbf{U}}}
\newcommand{\BR}{\ensuremath{\mathbf{R}}}
\newcommand{\BJ}{\ensuremath{\mathbf{J}}}
\newcommand{\Balpha}{\ensuremath{\boldsymbol{\alpha}}}
\newcommand{\Bbeta}{\ensuremath{\boldsymbol{\beta}}}
\newcommand{\Bphi}{\ensuremath{\boldsymbol{\phi}}}
\newcommand{\Btau}{\ensuremath{\boldsymbol{\tau}}}
\newcommand{\Bpi}{\ensuremath{\boldsymbol{\pi}}}
\newcommand{\Bxi}{\ensuremath{\boldsymbol{\xi}}}
\newcommand{\Br}{\ensuremath{\mathbf{r}}}
\newcommand{\Bx}{\ensuremath{\mathbf{x}}}
\newcommand{\Bq}{\ensuremath{\mathbf{q}}}
\newcommand{\Bp}{\ensuremath{\mathbf{p}}}
\newcommand{\Bv}{\ensuremath{\mathbf{v}}}
\newcommand{\Bh}{\ensuremath{\mathbf{h}}}
\newcommand{\BH}{\ensuremath{\mathbf{H}}}
\newcommand{\Ba}{\ensuremath{\boldsymbol{a}}}
\newcommand{\BPhi}{\ensuremath{\boldsymbol{\Phi}}}
\newcommand{\Bell}{\ensuremath{\boldsymbol{\ell}}}
\newcommand{\Bsigma}{\ensuremath{\boldsymbol{\sigma}}}

%other letters &operators
\newcommand{\Z}{\ensuremath{\mathds{Z}}}
\newcommand{\NN}{\ensuremath{\mathds{N}}}
\newcommand{\R}{\ensuremath{\mathds{R}}}
\newcommand{\SU}{SU(2)}
\newcommand{\su}{{su(2)}}
\newcommand{\e}{\mathrm{e}}
\newcommand{\rt}{\tilde{r}}
\newcommand{\di}{\mathrm{d}}
\newcommand{\dd}{\mathrm{d}}
\newcommand{\ii}{\mathrm{i}}
\newcommand{\CH}{\ensuremath{\mathcal{H}}}
\newcommand{\CL}{\ensuremath{\mathcal{L}}}
\newcommand{\CB}{\ensuremath{\mathcal{B}}}
\newcommand{\CM}{\ensuremath{\mathcal{M}}}
\newcommand{\CG}{\ensuremath{\mathcal{G}}}
\newcommand{\CE}{\ensuremath{\mathcal{E}}}
\newcommand{\CA}{\ensuremath{\mathcal{A}}}
\newcommand{\CF}{\ensuremath{\mathcal{F}}}
\newcommand{\be}{\begin{equation}}
\newcommand{\bea}{\begin{eqnarray}}
\newcommand{\tr}{\mbox{tr}}
\newcommand{\la}{\lambda}
\newcommand{\ta}{\theta}
\newcommand{\f}{\phi}
\newcommand{\vf}{\varphi}
\newcommand{\ka}{\kappa}
\newcommand{\al}{\alpha}
\newcommand{\ga}{\gamma}
\newcommand{\de}{\delta}
\newcommand{\si}{\sigma}
\newcommand{\bomega}{\mbox{\boldmath $\omega$}}
\newcommand{\bsi}{\mbox{\boldmath $\sigma$}}
\newcommand{\bchi}{\mbox{\boldmath $\chi$}}
\newcommand{\bal}{\mbox{\boldmath $\alpha$}}
\newcommand{\bpsi}{\mbox{\boldmath $\psi$}}
\newcommand{\brho}{\mbox{\boldmath $\varrho$}}
\newcommand{\beps}{\mbox{\boldmath $\varepsilon$}}
\newcommand{\bxi}{\mbox{\boldmath $\xi$}}
\newcommand{\bbeta}{\mbox{\boldmath $\beta$}}
\newcommand{\ee}{\end{equation}}
\newcommand{\pa}{\partial}
\newcommand{\Om}{\Omega}
\newcommand{\vep}{\varepsilon}
\newcommand{\bfph}{{\bf \phi}}
\newcommand{\eea}{\end{eqnarray}}

\title{Hopfion canonical quantization}

\author{
{\large A.~Acus}$^{\dagger}$, {\large A.~Halavanau}$^{\star}$,
 {\large E.~Norvai\v{s}as}$^{\dagger}$
and {\large Ya. Shnir}$^{\star \ddagger}$ \\ \\
\\ $^{\dagger}${\small Vilnius University, Institute of Theoretical Physics and Astronomy}
%\\ $^{\ddagger}${\small Department of Mathematical Physics, National
%University of Ireland Maynooth,}
\\ {\small Go\v{s}tauto 12, Vilnius 01108, Lithuania}
\\ $^{\star}${\small Department of Theoretical Physics and Astrophysics, BSU, Minsk, Belarus}
\\ $^{\ddagger}${\small Institute of Physics, Carl von Ossietzky University Oldenburg, Germany}
} \maketitle

\begin{abstract}
We study the effect of the canonical quantization of the rotational mode of the charge $Q=1$ and
$Q=2$ spinning Hopfions. The axially-symmetric solutions are constructed numerically, it is shown the
quantum corrections to the mass of the configurations are relatively large.

\end{abstract}

%\begin{keywords}
%\end{keywords}
%\classification{42.65.Hw, 42.65.Wi} % PACS numbers go here

\section{Introduction}

Since the early 1960s, the topological solitons  have been intensively
studied in many different frameworks. These localized regular field configuration are rather a common presence
in non-linear theories, they arise as solutions of the corresponding field equations in various
space-time dimensions. Examples in 3+1 dimensions include well known solutions
of the Skyrme model \cite{Skyrme:1961vq}, monopoles in Yang-Mills-Higgs theory \cite{Hooft-Polyakov}
and the solitons in the Faddeev-Skyrme model \cite{Faddeev-Hiemi},\cite{Gladikowski:1996mb}.

Though the structure of the Lagrangian of the Faddeev-Skyrme model is exactly the same as Skyrme
theory, the topological properties
of these models are very different, while in the former model the $O(4)$ scalar field is the map ${S}^3 \to {S}^3$, the triplet of
the Faddeev-Skyrme fields is the first Hopf map ${S}^3 \to {S}^2$. It was shown that solutions of the latter model should be not just
closed  flux-tubes of the fields but knotted field configurations \cite{Battye1998}. Consequent analysis revealed a very rich
structure of the Hopfion spectrum \cite{Hietarinta2000,Sutcliffe:2007ui}. A number of different models which describe topologically
stable knots associated with the first Hopf map $S^3 \to S^2$ are known in different contexts. It was argued, for example, that a
system of two coupled  Bose condensates may support Hopfion-like solutions \cite{Babaev}, or that glueball configurations in QCD may
be treated as Hopfions \cite{Kondo:2006sa}.

One of the reasons for the interest in Skyrme model is related with the suggestion that,
in the limit of large number of quark
colours there is  a
relation between this model and the low-energy QCD
with an identification between topological charge of the Skyrmion and baryon number \cite{Adkins:1983hy,Adkins:1983ya}.
This approach involves a study of spinning Skyrmions and semiclassical quantization of the rotational collective
coordinates as a rigid body.

The classical Skyrmion is usually quantized within the Bohr-Sommerfeld framework by requiring the angular momentum to
be quantized, i.e., the quantum excitations correspond to a spinning Skyrmion with a particular rotation frequency.
In the recent paper \cite{Battye:2005nx} an axially symmetric ansatz was used to allow the spinning Skyrmion to deform.
Furthermore, it was suggested to treat the Skyrme model quantum mechanically, i.e., apply the canonical quantization
of the collective
coordinates of the soliton solution to take into account quantum mass corrections
\cite{Fujii:1986wt}-\cite{Jurciukonis:2005em}.
It turns out the correction decreases the mass
of the spinning Skyrmion, so one can expect similar effect in the Faddeev-Skyrme model.

Similarity between the Lagrangians of the Faddeev-Skyrme and Skyrme models suggests to take into account
(iso)rotational collective degrees of freedom of the Hopfions whose excitation may
contribute to the kinetic energy of the configuration and strongly affect other properties of the spinning Hopfions
\cite{JHSS}.
An obviously relevant generalization then is related with canonical quantization of the rotational excitations.

Though the spinning Hopfions were considered in early paper \cite{Gladikowski:1996mb}, a systematic study
of their properties was not performed yet. One of the reason of that is that consistent consideration of the soliton
solution of the Faddeev-Skyrme model is related with rather complicated task of full 3d numerical simulations
\cite{Hietarinta2000,Sutcliffe:2007ui}. However this task becomes much simpler if we restrict our consideration to the
case of the axially symmetric Hopfions of charge 1 and 2.
In this Letter we are mainly concerned with canonical quantization of the rotational collective coordinates of these
Hopfions.

\section{The model}
Let us begin with a brief review of the  Faddeev-Skyrme model in 3+1 dimensions which is the $O(3)$-sigma model modified
by including a quartic term:
\be \label{model}
{\cal L} = \frac{1}{32\pi^2}\left(\partial_\mu \phi^a \partial^\mu \phi^a -
\frac{\kappa}{4}(\varepsilon_{abc}\phi^a\partial_\mu \phi^b\partial_\nu \phi^c)^2 - \mu^2 [1-(\phi^3)^2] \right)
\ee
Here $\phi^a = (\phi^1, \phi^2,\phi^3)$ denotes a triplet of scalar real
fields which satisfy the constraint $|\phi^a|^2=1$.
For finite energy solutions the field $\phi^a$ must tend to a
constant value at spatial infinity, which we select to be $\phi^a(\infty) = (0,0,1)$.
This allows a one-point compactification $R^3 \sim S^3$, thus
topologically the field is the map
$\phi({\bf r}):\mathbb{R}^3 \to S^2$ characterized by the Hopf invariant $Q = \pi_3(S^2) = \mathbb{Z}$ and
$\mu^2 [1-(\phi^3)^2]$ is the "pion" mass term which is included to stabilize the spinning soliton.
Note that our choice for this term is a bit different from the usual mass term in the conventional Skyrme model
(i.e., $\mu^2(1-\phi^3)$ ) since for the fields on the unit sphere it seems to be more convenient to perform
numerical calculations.

The energy of the Faddeev-Skyrme model is bound from below by the Vakulenko-Kapitansky inequality \cite{VK}
$E\geq {\rm const} |Q|^\frac{3}{4}$. In the classical case one can rescale the Lagrangian (\ref{model})
to absorb the coupling $\kappa$ into the rescaled mass constant,
however consequent canonical quantization of the spinning
Hopfion does not allow us to scale this constant away.

For the lowest two values of the Hopf charge $Q=1,2$ the Hopfion solutions  can be constructed on the
axially symmetric ansatz \cite{Gladikowski:1996mb} parametrised by two functions $f=f(r,\theta)$ and $g=g(r,\theta)$ of $r,\theta$ as a triplet of the scalar fields in circular coordinate system
\begin{eqnarray}
\label{ansatz}
\phi_+&=&-\frac{1}{\sqrt{2}}\sin f(r,\theta) \mathrm{e}^{\mathrm{i}(n\varphi -m g(r,\theta))},\nonumber\\
\phi_0&=&\cos f(r,\theta), \nonumber\\
\phi_{-}&=&\frac{1}{\sqrt{2}}\sin f(r,\theta) \mathrm{e}^{-\mathrm{i}(n\varphi -m g(r,\theta)).}
\end{eqnarray}
where $n,m \in \mathbb{Z}$.
An axially-symmetric configuration of
this type ${\cal A}_{m,n}$ has topological charge $Q = mn$, where the first subscript
labels the number of twists along the loop
and the second is the usual $O(3)$ sigma model winding number associated with the
map $S^2 \to S^2$, thus the ansatz (\ref{ansatz}) corresponds to the configurations ${\cal A}_{1,1}$ and
${\cal A}_{2,1}$.

Furthermore, one readily verifies
that the parametrization (\ref{ansatz}) is consistent, i.e. the complete set of the field equations, which
follows from the variation of the original action of the model (\ref{model}), is compatible with
two equations which follow from variation of the reduced action on ansatz (\ref{ansatz}).
However this trigonometric parametrization is not very convenient from the point of view of numerical
calculations because of the numerical errors which originate from the disagreement between
the boundary conditions on the angular-type function $g(r,\theta)$ on the $\rho$-axis and the boundary
points $r = 0,\infty$, respectively\footnote{Note that numerical difficulties of the same type are common in the
Skyrme model \cite{ST}.}.
Indeed, the reduced classical rescaled two-dimensional energy density functional, resulting from the
imposition of axial symmetry stated in ansatz (\ref{ansatz}), is given by
\be \label{redeng}
\begin{split}
{\cal M}(f,g)=&\frac{1}{32\pi^2 r^2}\biggl[
\frac{n^2\sin^2 f}{\sin^2 \theta}+\Bigl(\frac{\partial f}{\partial\theta}\Bigr)^2
+r^2 \Bigl(\frac{\partial f}{\partial r}\Bigr)^2 \\
&+m^2\sin^2 f \biggl(\Bigl(\frac{\partial g}{\partial\theta}\Bigr)^2+r^2 \Bigl(\frac{\partial g}{\partial r}\Bigr)^2\biggr)\\
&+\frac{\sin^2 f}{2 r^2} \biggl(\frac{n^2}{\sin^2 \theta}\biggl(\Bigl(\frac{\partial f}{\partial\theta}\Bigr)^2+r^2
\Bigl(\frac{\partial f}{\partial r}\Bigr)^2\biggr)\\
&+m^2 r^2 \Bigl(\frac{\partial f}{\partial r} \frac{\partial g}{\partial \theta}-\frac{\partial f}{\partial
\theta} \frac{\partial g}{\partial r}\Bigr)^2\biggr)+\mu^2r^2\sin^2 f \biggr]
\end{split}
\ee

\begin{figure}[hbt]
\lbfig{fig:1}
\begin{center}
(a)\hspace{-0.6cm}
\includegraphics[height=.27\textheight, angle =0]{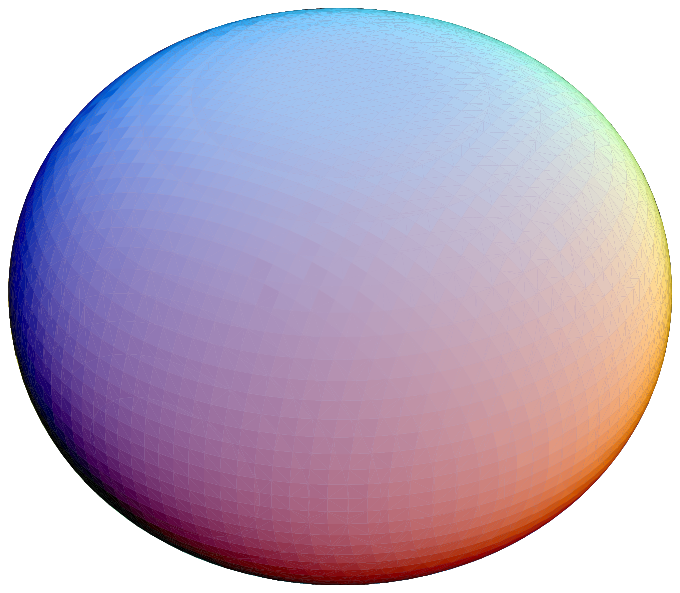}
\hspace{0.5cm} (b)\hspace{-0.6cm}
\includegraphics[height=.26\textheight, angle =0]{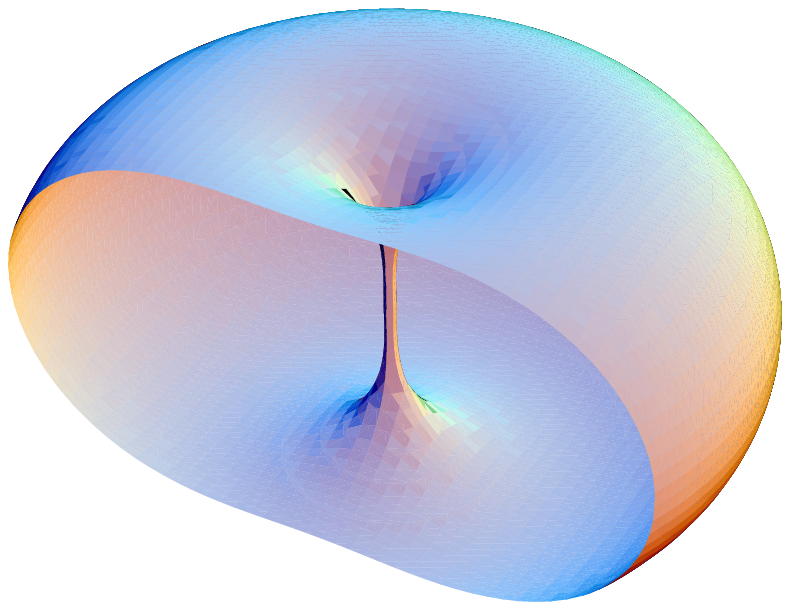}
\end{center}
\vspace{-0.5cm}
\caption{\small The energy isosurfaces for the charge 1 (a) and the charge 2 (b)
static Hopfions at $\mu^2=2$ and $\omega=0$.}
\end{figure}
The resulting system of the Euler-Lagrange equations can be solved when we impose
the boundary conditions  such that the resulting field configuration will be regular on the symmetry axis, at the origin and
on the spatial asymptotic.

The charge $Q=1$  ${\cal A}_{1,1}$ configuration possesses the maximum of the energy density at the origin, the energy
density isosurfaces are squashed spheres as seen in Fig.\ref{fig:1}, left. The charge $Q=2$  ${\cal A}_{2,1}$ solutions
have toroidal structure( see Fig.\ref{fig:1}, right). Inclusion of the mass term increases the
attraction in the system, the total energy of the massive Hopfion increases monotonically as mass parameter $\mu$
increases \cite{Foster:2010zb}.

The residual $O(2)$ global symmetry of the ansatz (\ref{ansatz}) with respect to the rotations around the
third axis in the internal space allows us to consider the stationary spinning classical Hopfions
\be \label{spinning}
\phi_+ \to \phi_+\mathrm{e}^{\mathrm{i}\omega t}; \qquad \phi_- \to \phi_-\mathrm{e}^{-\mathrm{i}\omega t}
\ee
Here, to secure stability of the configuration with respect to radiation, the rotation
frequency $\omega$ is a parameter restricted to the interval
\be
0 \le \omega < \mu
\ee
Substituting this ansatz into the lagrangian (\ref{model}) gives
\be
\label{rot-lag}
L = -M + \frac{\omega^2 \Lambda}{2}
\ee
where $M$ is the static energy of the Hopfion and $\Lambda$ is the moment of inertia
\be
\label{moment-inertia}
\Lambda =\frac{1}{16\pi}\int \sin \theta \mathrm{d}r \mathrm{d}\theta
\biggl[ \sin^2 f \biggl( 2 r^2  +   \Bigl(\frac{\partial f}{\partial\theta}\Bigr)^2+
r^2 \Bigl(\frac{\partial f}{\partial r}\Bigr)^2\biggr)
\biggr] \, ,
\ee
and the conserved quantity is the classical spin of the rotating configuration $J=\omega \Lambda$.

Note that the structure of the expression for the density of the moment of inertia (\ref{moment-inertia})
in the rigid body
approximation  does not
depend on the phase function $g(r,\theta)$. However the function $f(r,\theta)$ is angle dependent.

The mass of the static Hopfion as a function of the parameter $\mu$ is presented in
Fig. \ref{fig:3}, as $\mu=0$ the corresponding values of the Hopfion mass and the moment of inertia are
$M_{Q=1}=1.23, ~~\Lambda_{Q=1}=0.63$ and $M_{Q=2}=1.97, ~~\Lambda_{Q=2}=0.41$

\begin{figure}[hbt]
\lbfig{fig:3}
\begin{center}
\includegraphics[height=.38\textheight, angle =270]{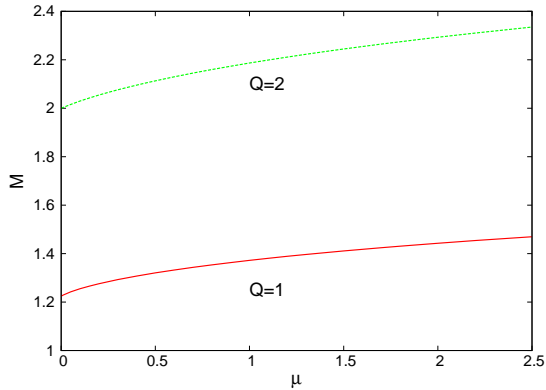}
\end{center}
\caption{\small The static energy of the axially-symmetric ${\cal A}_{1,1}, {\cal A}_{2,1}$ Hopfions as
function of the mass parameter $\mu$ at $\omega=0$.}
\end{figure}

As the angular velocity $\omega$ increases,  the total energy of the spinning configuration
as well as the moment of inertia and the angular momentum are increasing monotonically \cite{JHSS}.
Investigation of the energy density distribution reveal very interesting picture, as $\omega$ increases
a hollow circular tube is formed inside the Hopfion energy shell,
both for the charge 1 and charge 2 as shown in Figs.\ref{fig:4}. The moment
of inertia of the configuration diverges as $\omega \to \mu$.
\begin{figure}[hbt]
\lbfig{fig:4}
\begin{center}
(a)\hspace{-0.6cm}
\includegraphics[height=.23\textheight, angle =0]{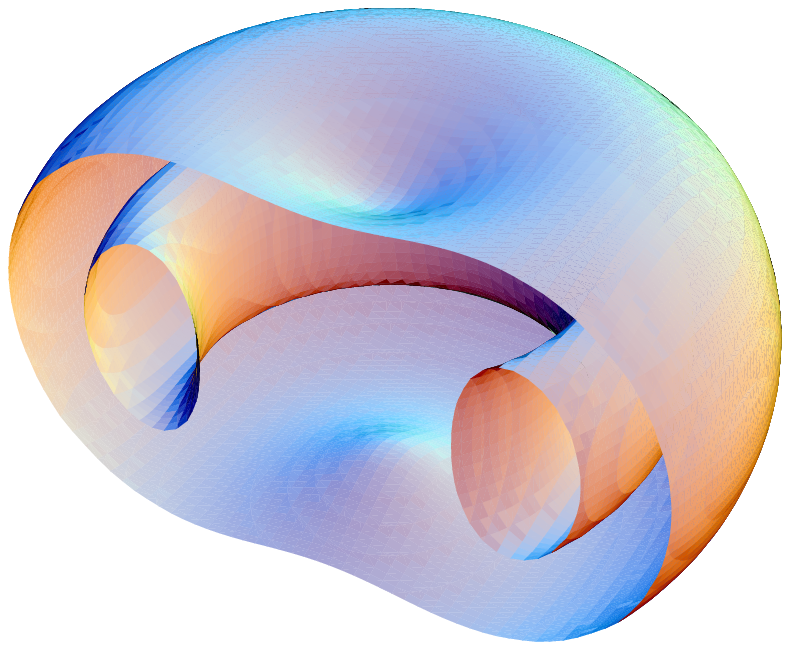}
\hspace{0.5cm} (b)\hspace{-0.6cm}
\includegraphics[height=.20\textheight, angle =0]{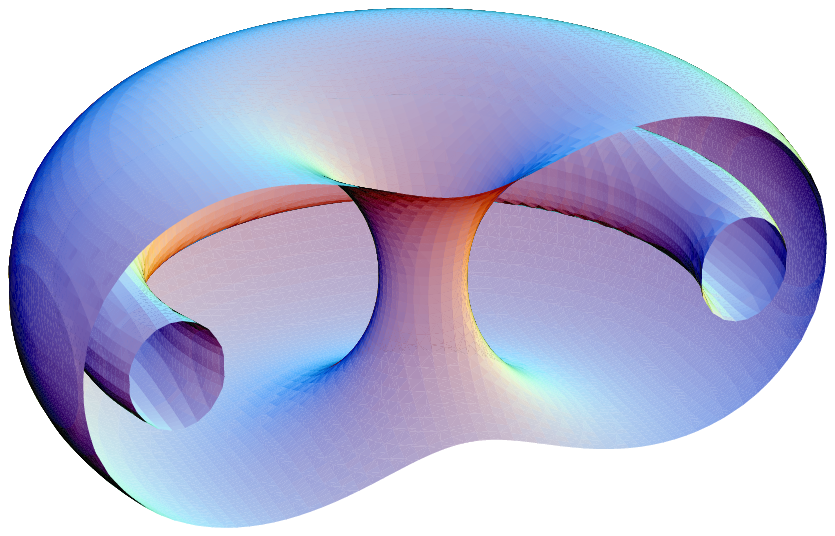}
\end{center}
\vspace{-0.5cm}
\caption{\small The energy isosurfaces of the ${\cal A}_{1,1}$ (a) and the ${\cal A}_{2,1}$ (b)
spinning Hopfions at $\omega^2 \sim \mu^2=2$.}
\end{figure}

The classical spinning Hopfion can be quantized within the Bohr-Sommerfield scheme by requiring the spin
to be quantized as $J^2 = j(j +1)$, where $j$ is the rotational quantum number taking half-integer values
\cite{Gladikowski:1996mb,Su:2001zw}.
The difference between our approach, where rotation occurs only around $z$ axis and therefore is characterized by of U(1) representations (i.e. takes only integer values), and the discussion presented
in the paper \cite{Su:2001zw} in that
in the latter case the charge $Q=1$  ${\cal A}_{1,1}$ configuration was considered by
a analogy with the case of the spinning Skyrmion where the usual hedgehog ansatz
$U=\exp \left(\mathrm{i} F(r) (\hat n^a \cdot \tau_a) \right)$
with a single radially dependent profile function $f(r)$ was implemented instead of the parametrization (\ref{ansatz}).
The relation between these two parametrizations can be explicitly written as
\begin{eqnarray}
\label{ansatzSU}
\phi_+&=&\sqrt{2}\sin F(r) \sin\theta\mathrm{e}^{-\mathrm{i}\varphi}\Bigl(\sin F(r) \cos\theta -\mathrm{i}\cos F(r)\Bigr),\nonumber\\
\phi_0&=&\cos (2\theta) \sin^2 F(r)  +\cos^2 F(r), \nonumber \\
\phi_{-}&=&-\sqrt{2}\sin F(r) \sin\theta\mathrm{e}^{\mathrm{i}\varphi}\Bigl(\sin F(r) \cos\theta +\mathrm{i}\cos F(r)\Bigr)
\end{eqnarray}
The functions $f(r,\theta)$ and $g(r,\theta)$ which parametrize the axially-symmetric ansatz (\ref{ansatz})
are related to the approximation by radial function $F(r)$ of \cite{Su:2001zw} as
\begin{eqnarray}
\label{suExplicit}
\cos f(r,\theta)&=&\cos (2\theta) \sin^2 F(r)  +\cos^2 F(r),\\
\tan g(r,\theta) &=&\frac{\cos F(r)}{\sin F(r) \cos\theta}.
\end{eqnarray}
Surprisingly, the hedgehog parametrization works extremely well for the minimal energy ${\cal A}_{1,1}$ configuration. It was pointed out
also by Ward \cite{Ward} who used the stereographic parametrization of the ${\cal A}_{1,1}$ and ${\cal A}_{2,1}$ Hopfions
in terms of the single radial-dependent
function ${\cal F}(r)$. For the former case this parametrisation is:
\begin{equation}
\label{wardImplicit}
W=\frac{x+\mathrm{i} y}{z-\mathrm{i}{\cal{F}}}=\frac{\sin\theta}{\sqrt{\cos^2\theta +\frac{{\cal{F}}^2}{r^2}}}\mathrm{e}^{\mathrm{i}\bigl(\varphi+\mathrm{arctg} \frac{{\cal{F}}}{r\cos\theta}\bigr)}
\end{equation}
The relation to the ansatz (\ref{ansatz}) is given by the expression
\begin{equation}
\label{wardCircular}
\phi_a =\frac{\sqrt{2}}{1+|W|^2} \bigl(-W,\frac{1}{\sqrt{2}}(1-|W|^2),\bar W\bigr) \, ,
\end{equation}
thus, we can represent the profile  functions  $f(r,\theta)$ and $g(r,\theta)$ as
\begin{eqnarray}
\label{wardExplicit}
\cos f(r,\theta)&=&\frac{r^2\cos 2\theta+{\cal{F}}^2}{r^2+{\cal{F}}^2},\\
\mathop{\mathrm{tg}}\, g(r,\theta) &=&-\frac{{\cal{F}}}{r\cos\theta}.
\end{eqnarray}
Finally, note that these two radial functions $F(r)$ and ${\cal F}(r)$ which are used in the  parametrizations
(\ref{ansatzSU}) and (\ref{wardImplicit}), respectively, are related as
\begin{equation}
\label{wardSu}
{\cal{F}} =-2\mathop{\mathrm{ctg}}\, F(r)
\end{equation}

Thus we will revisit the problem of the canonical quantization of the Hopfion using approach previously discussed in
\cite{Fujii:1986fu}-\cite{Jurciukonis:2005em}. For the sake of simplicity here we restrict  our analyse to the case of
the axially-symmetric configurations ${\cal A}_{1,1},{\cal A}_{2,1}$.

Similarity of the Lagrangian (\ref{model})
with the conventional Skyrme model suggests that in order
to apply the standard canonical quantization procedure it is
convenient to re-express the expression (\ref{model}) in terms of the hermitian matrix fields
\begin{equation}
\label{hopfAnz}
H=
\left(
\begin{array}{ll}
 \cos f(r,\theta) &\sin f(r,\theta) \mathrm{e}^{-\mathrm{i}(n\varphi -m g(r,\theta)) }  \\
 \sin f(r,\theta) \mathrm{e}^{\mathrm{i}(n\varphi -m g(r,\theta))} & -\cos f(r,\theta)
\end{array}
\right).
\end{equation}
which parametrises the Hopfion configuration. This matrix can be written compactly as
\begin{equation}
\label{hopfSum}
H=
2 \sum_a (-1)^a \tau_a \phi_{-a}; \qquad H\cdot H =\mathbf{1}
\end{equation}
where the usual algebra of the Pauli matrices $(\tau_+, \tau_0, \tau_-)$
yields
\begin{equation}
\label{tau1}
\tau_a \tau_b =\frac14 (-1)^a \delta_{a,-b}\mathbf{1} -\frac{1}{\sqrt{2}}
\left[
\begin{matrix}
1 & 1 &1\\
a & b & c
\end{matrix}
\right]\tau_c .
\end{equation}
Here the symbol in the square brackets is the $SU(2)$ Clebsh-Gordan coefficient.

In this notations the Lagrangian (\ref{model}) can be rewritten
as (the metric $\mathop{\mathrm{diag}}(1,-1,-1,-1)$ is explicitly assumed).
\begin{equation}\label{LagrangianDensity1}
{\cal L} =
\frac{1}{64\pi^2} \left( \mathop{\mathrm{Tr}} \partial_\mu H \partial^\mu H + \frac{\kappa}{16}
\mathop{\mathrm{Tr}}\bigl[ \partial_\mu H, \partial_\nu H \bigr]\bigl[ \partial^\mu H, \partial^\nu H \bigr]-
\frac{\mu^2}{2}\mathop{\mathrm{Tr}} \bigl(\mathbf{1}-4\tau_0 H\tau_0 H\bigr)\right).
\end{equation}

\section{Quantization. Momenta of inertia}
Similarity of the form of the Lagrangian (\ref{LagrangianDensity1}) with that of the
Skyrme model suggests that we can
quantize the rotational degrees of freedom of the axially-symmetric Hopfion
by wrapping the ansatz (\ref{hopfAnz}) with time-dependent unitary matrices $\BA\bigl(q(t)\bigr)$ \cite{Adkins:1983ya}
which rotates the configuration about the third axis:
\begin{equation}
\BU(q,f,g)=\BA\bigl(q(t)\bigr) H \BA^\dagger\bigl(q(t)\bigr).
\end{equation}
Thereafter the collective rotational degrees of freedom $q(t)$ are treated as quantum-mechanical variables, i.e.
the generalized  rotational coordinate $q(t)$ and velocity $\dot q(t)$ satisfy the commutation relations
\be
\label{com}
\bigl[\dot q, q\bigr]=\mathrm{i} f_{00}\,.
\ee
The explicit form of the constant $f_{00}$ will be completely determined by canonical
commutation relations between quantum coordinates
and momenta. As usual, to calculate the effective Lagrangian of the rotational
zero mode we have to evaluate the time derivative of the matrix
\begin{eqnarray}
\dot \BU=\dot \BA H \BA^\dagger - \BA H \BA^\dagger\dot \BA \BA^\dagger,\\
\nabla_k U = \BA \nabla_k H \BA^\dagger
\end{eqnarray}

Taking into account the commutation relation (\ref{com}) we obtain
\begin{equation}
\label{parametrization}
\BA(q)
=\exp\bigl(\mathrm{i}q \tau_0\bigr);\qquad \BA^\dagger\dot \BA=\mathrm{i} \tau_0 \dot q +\frac{\mathrm{i}}{8} f_{00}\mathbf{1} .
\end{equation}
Then, keeping only terms proportional to the square of the angular velocity the effective kinetic Lagrangian density can be written as
\begin{equation}
\CL_q \approx \frac{\sin^2 f}{64\pi^2 r^2}\dot q^2 \biggl(
2 r^2  + \Bigl(\frac{\partial f}{\partial\theta}\Bigr)^2
+r^2 \Bigl(\frac{\partial f}{\partial r}\Bigr)^2\biggr) \equiv \frac12\dot q^2 ~ g_{00}  \, .
\end{equation}
Utilizing the definition of the moment of inertia (\ref{moment-inertia}) we can
write
\begin{equation}
\label{quantum}
L_q = \frac{1}{2}\dot q^2 \int\mathrm{d}^3 r g_{00}=\frac{1}{2}\dot q^2 \Lambda
\end{equation}
Thought the expression (\ref{quantum}) coincides with its classical counterpart in (\ref{rot-lag}),
the corresponding quantum momentum is conjugated to the rotational
collective coordinate $q$ and it is defined as
\begin{equation}
\hat p=\frac{\partial L_q}{\partial \dot q}=\Lambda \dot q
\end{equation}
Thus, the canonical commutation relation
$\bigl[\hat p, q\bigr]=-\mathrm{i}$ allows us to define
$f_{00}=\frac{1}{\Lambda}$. We can also define the U(1) group generator which is the angular momentum operator
\begin{equation}
\hat \BJ=-\hat p =- \Lambda \dot q
\end{equation}
for eigenstates $|k\rangle = \mathrm{e}^{-\mathrm{i} k q} |0\rangle$ with integer eigenvalues $k=0,\pm 1, \pm 2,\ldots$.

We are now in position to evaluate the explicit form of the
quantum-mechanical Lagrangian of the Faddeev-Skyrme
model. Using expression \eqref{parametrization} we obtain:
\begin{equation}
\begin{split}
\CL_q&=\frac{1}{64\pi^2}\biggl(\mathop{\mathrm{Tr}}\dot \BU \dot \BU -\frac{1}{8} (-1)^a \mathop{\mathrm{Tr}}\BA
\bigl[\bigl[\BA^\dagger\dot \BA,H\bigr], \nabla_a H\bigr]\bigl[\bigl[\BA^\dagger\dot \BA,H\bigr], \nabla_{-a}
H\bigr]\BA^\dagger\biggr)\\
&=
\frac{\sin^2 f}{64\pi^2 }\left( \dot q^2 + \frac{1}{4\Lambda^2} \right)
\biggl(
2 +\frac{1} {r^2}\biggl(\Bigl(\frac{\partial f}{\partial\theta}\Bigr)^2+
\Bigl(\frac{\partial f}{\partial r}\Bigr)^2\biggl)\biggl)\\
&= \frac{\sin^2 f}{64\pi^2 \Lambda^2}\left( \hat \BJ^2 + \frac{1}{4} \right)
\biggl(
2 +\frac{1} {r^2}\biggl(\Bigl(\frac{\partial f}{\partial\theta}\Bigr)^2+
\Bigl(\frac{\partial f}{\partial r}\Bigr)^2\biggl)\biggl)
\end{split}
\end{equation}

The total effective Hamiltonian corresponds to the complete Lagrangian $L=L_\mathrm{cl}+L_q
$which includes both classical and quantum mechanical parts:
\begin{equation}
H=\frac12 \bigl\{\hat p, \dot q \bigr\}-L=
\frac{\hat \BJ^2}{2 \Lambda} -L_\mathrm{cl} +\Delta M
\end{equation}
Here the quantum mass correction $\Delta M$ appears when the canonical commutation relation
is taken into account:
\be \label{deltaM}
\Delta M = - \frac{1}{8 \cdot 16 \pi \Lambda^2} \int \sin \theta \mathrm{d}r \mathrm{d}\theta
\biggl[ \sin^2 f \biggl( 2 r^2  +  \Bigl(\frac{\partial f}{\partial\theta}\Bigr)^2+
r^2 \Bigl(\frac{\partial f}{\partial r}\Bigr)^2\biggr)
\biggr] = - \frac{1}{ 8 \Lambda}
\ee
where we used the definition (\ref{moment-inertia}). Note that an interesting peculiarity of the
integrand in (\ref{deltaM}) is that it exactly reproduces the structure of the density
of the moment of inertia (\ref{moment-inertia}), thus in the rigid body approximation we can
immediately evaluate the quantum corrections to the axially-symmetric configurations
${\cal A}_{1,1}, {\cal A}_{2,1}$ as
\be
\Delta M_{1,1} = -\frac{1}{ 8 \Lambda_{Q=1}} = - 0.20; \qquad
\Delta M_{2,1} = -\frac{1}{ 8 \Lambda_{Q=2}} = - 0.30
\ee
thus, for the configurations with topological charges $Q=1,2$ the quantum correction to the
Hopfion mass is negative and it is about $16 \%$ and $25 \%$ of the classical masses, respectively.

A more consistent treatment of the quantum correction to the Hopfion mass needs
minimization of the total energy functional
\begin{equation}
\begin{split}
\CH=-\CL_\mathrm{cl}+\frac{\sin^2 f}{32\pi^2r^2}\Bigl(\frac{\hat \BJ^2}{2 \Lambda^2}-\frac{1}{8 \Lambda^2}\Bigr)
\biggl(
2 r^2  +\Bigl(\frac{\partial f}{\partial\theta}\Bigr)^2
+r^2 \Bigl(\frac{\partial f}{\partial r}\Bigr)^2\biggl)
\end{split}
\end{equation}
Varying it we obtain rather cumbersome set of two coupled integro-differential equation for
functions $f(r,\theta)$ and $g(r,\theta)$ which then should be solved numerically. The results will be reported elsewhere.

\section*{Conclusion}

The main purpose of this letter was to present the scheme of the canonical quantization of the rotational mode of the
charge $Q=1$ and $Q=2$ spinning Hopfions and evaluate the quantum corretions to the mass of these axially-symmetric
configurations.
To this end we have used the technique described in \cite{Fujii:1986fu}-\cite{Jurciukonis:2005em} in the context of the
Skyrme model and Baby Skyrme model \cite{ANS}. The model is stabilised by additional coupling to a potential
(mass) term by analogy with
the Baby Skyrme model, this leads to appearance of the Yukawa-type exponential tail of the Hopfion fields.
The analysis of the  quantum corrections to the mass of the axially symmetric charge $Q=1,2$ solitons showed that,
like in the Skyrme model, the corrections are negative and relatively large.

It remains to systematically analyze the effect of quantization of the rotating Hopfions beyond the
usual Bohr–-Sommerfeld framework and the rigid body approximation we implemented in the present letter.
As a direction for future work, it would be interesting to study the effect of canonical quantisations of the
spinning knotted Hopfions, e.g. to consider how the shape of the celebrated $Q=7$ trefoil knot
configuration ${\cal K}_{3,2}$
will be affected by the quantum corrections or if the axial symmetry of the spinning charge $Q=3$ buckled configuration
will be restored.
Other buckling and twisting transmutations of the Hopfions which are
related with a change of the symmetry
of various spinning configurations of  higher Hopf degree are also possible, one can expect an axially symmetric
state may be the lowest energy state in this case.
This work is now in progress \cite{JHSS}.

%%%%%%%%%%%%%%%%%%%%%%%%%%%%%%%%%%%%%%%%%%%%%%%%%%%%
\section*{Acknowledgements}
%%%%%%%%%%%%%%%%%%%%%%%%%%%%%%%%%%%%%%%%%%%%%%%%%%%%
Ya.S. is very grateful to D.~Foster, D.~Harland, J.M.~Speight and P.~Sutcliffe
for many enlightening discussions. This work is supported by the A.~von Humboldt Foundation (Ya.S.).

\begin{small}

\end{small}

\end{document}